\documentclass[11pt,reprint,amsmath,amssymb,
 nofootinbib, aps,prl]{revtex4-2}
 
\usepackage{geometry}
\usepackage{amsmath,amsthm,amssymb, mathtools,mathrsfs}
\usepackage{physics}
\usepackage{amsfonts, latexsym, amssymb}
\usepackage{bbm, dsfont}
\usepackage{graphicx}
\usepackage{float}
\usepackage{url}
\usepackage{xcolor}
\usepackage[colorlinks, linkcolor=darkblue, citecolor=darkblue, urlcolor=darkblue, linktocpage]{hyperref} 
\usepackage{tikz}
\usetikzlibrary{decorations.pathmorphing,cd,decorations.markings,calc,fadings}
\hypersetup{
    colorlinks=true,
    linkcolor=blue,
    filecolor=magenta,      
    urlcolor=cyan,
}
\usepackage{enumerate}
\usepackage{bm}

\usepackage[export]{adjustbox}
\usepackage{cleveref}
\usepackage{appendix}
\usepackage{bbold}

\makeatletter
\def\l@subsection#1#2{}
\def\l@subsubsection#1#2{}
\makeatother

\definecolor{darkblue}{cmyk}{0.9,0.9,0,0}
\definecolor{darkgreen}{cmyk}{0.9,0,0.9,0}
\definecolor{blueblue}{cmyk}{0.73,0.28,0,0.5}
\definecolor{lightblue}{RGB}{55,171,200}
\definecolor{grey}{gray}{0.55}
\definecolor{pink}{cmyk}{0., 0.9859943977591037, 0.3571428571428571, 0.16000000000000003}
\definecolor{lightpink}{cmyk}{0., 0.5, 0.5, 0.}
\definecolor{lightgreen}{cmyk}{0.24175824175824182, 0., 0.9615384615384616, 0.28627450980392155}

\renewcommand{\d}{\mathrm{d}}

\newcommand{\ds}{\displaystyle}

\newcommand{\comment}[1]{}

\definecolor{darkgreen}{rgb}{0.1,0.6,0.1}

\newcommand{\rrangle}{\rangle\!\rangle}
\newcommand{\llangle}{\langle\!\langle}

\newcommand{\be}{\begin{equation}} \newcommand{\ee}{\end{equation}}
\newcommand{\ben}{\begin{equation*}} \newcommand{\een}{\end{equation*}}

\newcommand{\bea}{\begin{equation} \begin{aligned}} \newcommand{\eea}{\end{aligned} \end{equation}}

\newcommand{\cA}{\mathcal{A}}

\newcommand{\cC}{\mathcal{C}}

\newcommand{\cL}{\mathcal{L}}

\newcommand{\cM}{\mathcal{M}}

\newcommand{\bZ}{\mathbb{Z}}

\newcommand{\unit}{\mathbbm{1}}

\def\repa{\raise4pt\hbox{$\square$}\mkern-14mu\raise-4pt\hbox{$\square$}}
\def\repab{\overline{\raise4pt\hbox{$\square$}\mkern-14mu\raise-4pt\hbox{$\square$}\mkern-1mu}}

\def\({\left(}
\def\){\right)}
\def\[{\left[}
\def\]{\right]}
\def\<{\langle}
\def\>{\rangle}

\begin{document}

\title{Non-Invertible Symmetries, Anomalies and Scattering Amplitudes
}
\author{Christian Copetti$^{a}$, Lucía Córdova$^{b}$, Shota Komatsu$^{b}$}
\affiliation{${}^{a}$Mathematical Institute, University of Oxford, Woodstock Road, Oxford, OX2 6GG, United Kingdom}
\affiliation{${}^{b}$Department of Theoretical Physics, CERN, 1211 Meyrin, Switzerland}
\email{christian.copetti@maths.ox.ac.uk}
\email{lucia.gomez.cordova@cern.ch}
\email{shota.komatsu@cern.ch}


\begin{abstract}
    We show that crossing symmetry of S-matrices is modified in certain theories with non-invertible symmetries or anomalies. Focusing on integrable flows to gapped phases in two dimensions, we find that S-matrices derived previously from the bootstrap approach are incompatible with non-invertible symmetries along the flow. We present consistent alternatives, which however violate standard crossing symmetry and obey modified rules dictated by fusion categories. We extend these rules to theories with discrete anomalies.
\end{abstract}

\maketitle
\section{Introduction}
Symmetries and anomalies are fundamental concepts in theoretical physics. They constrain the dynamics of quantum field theory (QFT) by forbidding or guaranteeing certain interactions to be generated along the renormalization group (RG) flow and offer insights into infrared (IR) phases. They also play a crucial role in interpreting collider experiment data; symmetries provide organizing principles for particles and resonances while anomalies, like the Adler-Bell-Jackiw anomaly, are needed to explain  experimental observations such as $\pi^{0}\to\gamma\gamma$ decay. 

Recent theoretical progress involves refinement of these concepts, such as higher-form  or non-invertible symmetries \cite{Gaiotto:2014kfa, Frohlich:2004ef,  Fuchs:2007tx, Bhardwaj:2017xup, Chang:2018iay} and various non-perturbative anomalies \cite{Witten:1982fp,Dai:1994kq,Kapustin:2014lwa,Kapustin:2014zva,Witten:2016cio,Gaiotto:2017yup,Wang:2018qoy}. However, unlike their traditional counterparts, their implication on scattering amplitudes remains largely unexplored. This paper initiates such a study, with a primary focus on theories in two spacetime dimensions. The punchline of our analysis is simple yet striking:\vspace{5pt}\\
    {\it Crossing symmetry of S-matrices is modified in the presence of certain non-invertible symmetries or anomalies.}\vspace{5pt}\\
Our main examples are RG flows from unitary minimal models to gapped phases which preserve both integrability and non-invertible symmetries. We show that the basic four properties that one would postulate for these S-matrices --- unitarity, crossing symmetry, Yang-Baxter equation and non-invertible symmetries --- are mutually incompatible. We derive S-matrices satisfying all but crossing symmetry by modifying existing proposals in the literature. They obey modified crossing rules determined by fusion categories. These rules extend to theories with discrete anomalies as we see in the example of perturbed $SU(2)_1$ Wess-Zumino-Witten (WZW) model.
\section{Review of basic concepts}
\subsection{\texorpdfstring{Categorical symmetries in $1+1$ dimensions}{Categorical symmetries in 1+1 dimensions}}Non-invertible symmetry is a generalization of standard symmetries, which does not obey group multiplication laws. In $1+1$ dimensions, such symmetries are described by a set of topological line operators $\cL_a$ belonging to a (unitary) {\it fusion category} \cite{etingof2016tensor} $\cC$, as we review below. We will follow the conventions of \cite{Barkeshli:2014cna,Aasen:2020jwb}.  For physics applications, see e.g.~\cite{Frohlich:2006ch,Frohlich:2009gb,Carqueville:2012dk,Chang:2018iay,Komargodski:2020mxz,Aasen:2020jwb,Thorngren:2019iar,Thorngren:2021yso,Lootens:2021tet,Lin:2022dhv,Lin:2019kpn,Lin:2021udi,Lin:2023uvm}. 

\subsubsection{Fusion algebra and junction} Topological lines satisfy the fusion algebra,
 \be
\cL_{a}\cL_{b}=\sum_{\cL_c}N_{ab}{}^{c}\cL_c\,,
 \ee
with non-negative integers $N_{a  b}{}^c$. Not all  $\mathcal{L}_a$'s have inverse i.e.~$\mathcal{L}_a(\mathcal{L}_a)^{-1}=\mathbb{1}$, rendering them `non-invertible'. Line operators can form trivalent junctions. For a given set of lines ($\cL_{a,b,c}$), there can be multiple such choices, which form finite-dimensional vector spaces. We denote them by $V_{ab}{}^{c}$ or $V_{c}{}^{ab}$, depending on the orientation,
\be
\includegraphics[height=2.2cm,valign=c]{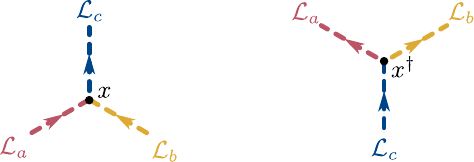} 
\ee
 and $\text{dim}(V_{a  b}{}^c) = N_{a  b}{}^c$. The junctions satisfy orthogonality and completeness relations,
 \begin{equation}
    \includegraphics[height=2.9cm,valign=c]{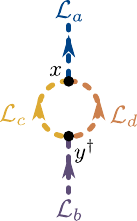}\, =\, \delta_{ab} \delta_{xy} \sqrt{\frac{\d_c \d_d}{\d_a}}\; \includegraphics[height=2.9cm,valign=c]{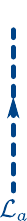}
    \label{eq:Lortho}
 \end{equation}
and
 \begin{equation}
     \includegraphics[height=2.9cm,valign=c]{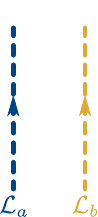} = \, \sum_{\cL_c , \, x} \, \sqrt{\frac{\d_c}{\d_a \, \d_b}}\, \includegraphics[height=2.9cm,valign=c]{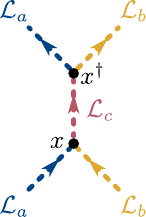}
     \label{eq:Lcomp}
 \end{equation}
Here $\d_a$ is the {\it quantum dimension}, defined by the expectation value\footnote{Precisely speaking, $\langle \mathcal{L}\rangle $ on a plane can differ from the quantum dimension by a phase due to the isotopy anomaly \cite{Chang:2018iay}. For instance, $\langle \mathcal{L}_{\mathbb{Z}_2}\rangle=-1$ if $\mathbb{Z}_2$ is anomalous.} of a closed loop of $\mathcal{L}_a$:
\be\label{eq:Lqd}
\langle \mathcal{L}_a\rangle=\includegraphics[height=1.2cm,valign=c]{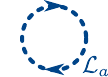}=\d_a \,.
\ee
Below we omit the junction labels ($x$ and $x^{\dagger}$) since $\text{dim}(V_{a  b}{}^c) =$ $0$ or $1$ in our examples. We also omit the arrows as we will discuss unoriented lines for which $\bar{\mathcal L}=\mathcal L$. 

\subsubsection{F-symbols and tetrahedral symbols} 
Also important in the fusion category are the {\it F-symbols}, which relate different networks of lines:
\begin{equation}
    \includegraphics[height=2.4cm,valign=c]{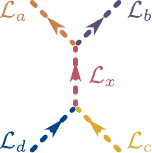} =\, \sum\limits_y {F^{abc}_d}_{xy}
    \includegraphics[width=2.4cm,valign=c]{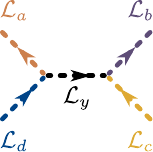}
    \label{eq:Fsymb}
\end{equation}

They satisfy a consistency condition known as the pentagon identity \cite{Moore:1988qv,etingof2016tensor}.
In actual computations, it is convenient to use the {\it tetrahedral symbols} defined by\footnote{Explicit expressions for these symbols in our examples are given in Supplemental Material.}
\be\label{eq:Ftetra}
\begin{bmatrix}
a & b & x \\ c & d & y     
\end{bmatrix} = \frac{1}{\sqrt{\d_x \d_y}} \, {F^{a b c}_d}_{x y} \,.
\ee
$F$-symbols encode the 't Hooft anomalies of a symmetry $\cC$. For a discrete group $G$, admissible $F$-symbols are given by cohomology classes $\omega\in H^{3}(G,U(1))$. A nontrivial $\omega$ signals an 't Hooft anomaly. The simplest example is $G=\mathbb{Z}_2$ with generator $\eta$, for which the $F$-symbol is a sign $\epsilon=\pm 1$:
\begin{equation}
\label{eq:Z2Fsymb}    \includegraphics[height=2.2cm,valign=c]{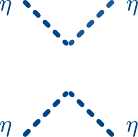} =\, \epsilon\;
    \includegraphics[width=2.2cm,valign=c]{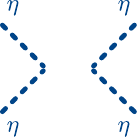}
\end{equation}
The symmetry is anomalous iff $\epsilon=-1$. In general, (non-invertible) symmetries are anomalous if there is no \emph{fiber functor} i.e.~a $\cC$ symmetric invertible topological QFT (TQFT) \cite{Thorngren:2019iar}.
The anomalies of non-invertible symmetries have been studied in e.g. \cite{Chang:2018iay,Kaidi:2023maf,Zhang:2023wlu,Cordova:2023bja,Antinucci:2023ezl}.

\subsubsection{Symmetry constraints and action on vacua}Non-invertible symmetries can constrain the number of vacua in the IR. In a $\mathcal{C}$-symmetric RG flow to a gapped phase, these must form a {\it module category} over $\mathcal{C}$ (see e.g.~\cite{Thorngren:2019iar,Huang:2021zvu,Bhardwaj:2023idu}), which generalizes the notion of group representations to fusion categories. Concretely, the action on the vacua $|i\rangle$ on a (large) circle reads
\be
\cL_a | i \rangle = \sum_j \, {n_a}_i{}^j  |j \rangle \, .
\ee
where ${n_a}_i{}^j$ provide a representation of the fusion algebra:
\be
\sum_{j}   {n_a}_i{}^j \, {n_b}_j{}^k = \sum_{c} N_{ab}{}^c {n_{c}}_i{}^k \, .
\ee
In addition, ${n_a}_i{}^j$ need to be non-negative integers; this can be shown by relating them to the ground-state degeneracy on a strip\footnote{Here we consider an infinitely large cylinder in order to project to the ground states.}:
\be
{n_{a}}_i{}^{j}=\includegraphics[width=2.0cm,valign=c]{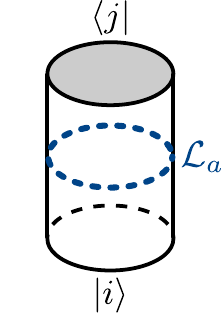}=\;\includegraphics[height=2.0cm,valign=c]{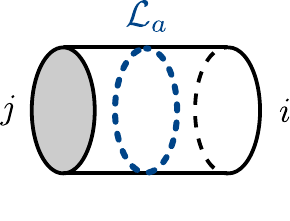}
\ee
A complete description of the module structure can be found e.g. in \cite{etingof2016tensor,Huang:2021zvu}.
Unlike standard (non-anomalous) symmetries, such representations are highly constrained, often lacking the trivial representation and thus forbidding a single vacuum in the IR.\footnote{A 'trivial' representation, if it exists, corresponds to the aforementioned fiber functor.} A necessary (but not sufficient) condition for a single vacuum is $\d_a\in \mathbb{N}_{>0}$ for any $a$, since ${n_{a}}_{0}{}^{0}=\d_a$ when the vacuum is unique. In our examples, this condition is violated and hence the IR has multiple vacua. 

Of particular importance is
the \emph{regular} representation; the largest irreducible representation corresponding to the choice $\{ i \} = \{ a \}$ and
$
n_{ab}{}^c = N_{ab}{}^c \, 
$. In such cases, all the vacua can be obtained from the `identity vacuum' $|0\rangle$ by acting $\mathcal{L}_a$:
\be\label{eq:genvac}
\mathcal{L}_a|0\rangle=|a\rangle\,.
\ee
Physically, this encodes the spontaneous symmetry breaking of $\cC$, realized for example by $\phi_{1,3}$-deformations of minimal models discussed below.

\subsection{\texorpdfstring{Integrable scattering in $1+1$ dimensions}{Integrable scattering in 1+1 dimensions}}
{\it Integrable QFTs} in $1+1$ dimensions offer rare instances of interacting QFTs for which exact computations are possible \cite{Zamolodchikov:1977nu,Zamolodchikov:1978xm}. This is due to {\it factorized scattering} --- a property that all scattering processes can be expressed as a sequence of two-body scatterings --- which in turn is a result of higher spin conserved currents \cite{Shankar:1977cm,Parke:1980ki}.
\subsubsection{S-matrix axioms}
Consider $2\to2$ S-matrices of identical mass $m$ particles. They depend on the center of mass energy $s=(p_1+p_2)^2=4m^4\cosh^2(\theta/2)$, where $\theta$ is the rapidity difference $\theta\equiv \theta_{12}=\theta_1-\theta_2$. As our focus will be on S-matrices of kinks interpolating between different vacua (denoted by $a$-$d$ below), we denote them by
\begin{equation}
    S^{ab}_{dc} (\theta)=\includegraphics[height=2cm,valign=c]{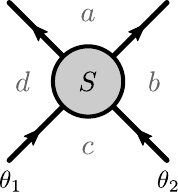}
\end{equation}

{\it Crossing symmetry} exchanges in and out particles. In the figure above, it amounts to changing the time direction from going upward to going rightward and corresponds to the following relation
\begin{equation}\label{eq:Scross0}
    S^{ab}_{dc} (\theta) = \includegraphics[height=2cm,valign=c]{figures/Stheta.pdf} =\includegraphics[height=2cm,valign=c]{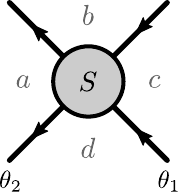} = S^{bc}_{ad} (i\pi-\theta)\,.
\end{equation}
In integrable QFTs, the S-matrices obey further constraints: {\it Unitarity} amounts to the equality 
\begin{equation}\label{eq:Sunit}
    \sum_e S^{eb}_{dc} (\theta) S^{ab}_{de} (-\theta)= \includegraphics[height=3.2cm,valign=c]{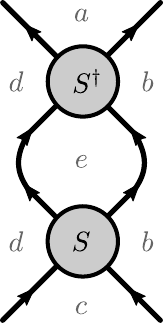} = \delta_{ac} \;\;\includegraphics[height=3.2cm,valign=c]{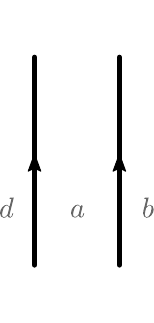} 
\end{equation}
while factorized scattering implies that S-matrices obey the {\it Yang-Baxter equations}
\begin{gather}
    \resizebox{\hsize}{!}{$
   \sum\limits_g S^{gd}_{fe}(\theta_{12}) S^{bc}_{gd}(\theta_{13}) S^{ab}_{fg}(\theta_{23}) =
   \sum\limits_g S^{gc}_{ed}(\theta_{23}) S^{ag}_{fe}(\theta_{13}) S^{bc}_{ag}(\theta_{12})
   $}\nonumber\\
   \sum\limits_g\;
   \includegraphics[height=2.7cm,valign=c]{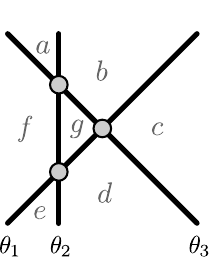} 
   = \sum\limits_g\;
   \includegraphics[height=2.7cm,valign=c]{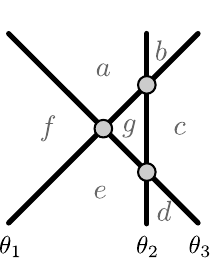} 
   \label{eq:SYB}
\end{gather}

\subsection{Minimal models and deformations}Canonical examples of QFTs with non-invertible symmetries are diagonal unitary minimal models $\mathcal{M}_n=\mathcal{M}_{n+1,n}$ and their relevant deformations. As is well-known \cite{DiFrancesco:1997nk}, $\mathcal{M}_n$ has a set  of primary fields $\phi_{r,s}$ ($1\leq r\leq n-1$, $1\leq s\leq n$) subject to the identification $\phi_{r,s}=\phi_{n-r,n+1-s}$. It also hosts topological lines $\mathcal{L}_{r,s}$ \cite{Petkova:2000ip,Chang:2018iay} whose fusion algebra coincides with that of primary fields. For instance, the Ising model $\mathcal{M}_3$ has three topological lines $1,\mathcal{N},\eta$, corresponding to $\phi_{1,1},\phi_{1,2},\phi_{1,3}$, which satisfy
\be
\eta^2=1\,,\quad \mathcal{N}\eta=\eta\mathcal{N}=\mathcal{N}\,,\quad \mathcal{N}^2=1+\eta\,.
\ee

The symmetry of $\mathcal{M}_n$ contains a simple subcategory $\mathcal{A}_{n-1}=\{\mathcal{L}_{r,1}\,,\, 1\leq r\leq n-1\}$. 
Its fusion algebra coincides with that of $SU(2)_{n-2}$ although the $F$-symbols are different.
\subsubsection{\texorpdfstring{$\phi_{1,3}$-deformation}{phi13-deformation}}Relevant deformations of $\mathcal{M}_n$ by $\phi_{1,3}$ ($\mathcal{M}_n+\lambda\int \phi_{1,3}$) have several interesting properties:
\begin{itemize}
\item Depending on the sign of the deformation, they either flow to $\mathcal{M}_{n-1}$ ($\lambda>0$) or to a gapped phase with $n-1$ degenerate vacua ($\lambda <0$).
\item They preserve integrability \cite{Zamolodchikov:1989hfa,Zamolodchikov:1991vx}.
\item They also preserve a non-invertible $\mathcal{A}_{n-1}$ symmetry. See \cite{Chang:2018iay,Kikuchi:2021qxz} and Supplemental Material for details. The vacua in the gapped phase transform as the regular representation of $\mathcal{A}_{n-1}$. 
\end{itemize}
\subsubsection{Other deformations} Deformations by $\phi_{1,2}$ or $\phi_{2,1}$ also preserve integrability \cite{Zamolodchikov:1989hfa,Zamolodchikov:1990xc,Smirnov:1991uw,Colomo:1991gw}. They flow to gapped phases irrespective of the sign of the deformation and preserve the subcategories generated by $\cL_{3,1}$ and $ \cL_{1,3}$ respectively.
Later we study in detail the $\phi_{2,1}$-deformation of the tricritical Ising model ($\mathcal{M}_4$).

\section{S-matrices with modified crossing}
We now discuss examples with modified crossing relation. We determine exact S-matrices by requiring consistency with symmetries and anomalies. The results differ from those in the literature and satisfy modified crossing rules. A physical derivation of these rules will be presented in section \ref{sec:modified}.

\subsection{Gapped flows from minimal models}
\subsubsection{Conjecture in the literature}
The first example is the  $\phi_{1,3}$-deformation of $\mathcal{M}_n$ to a gapped phase. The spectrum of this flow consists of kinks interpolating between neighboring vacua, all with identical mass $m$. S-matrices for these kinks were conjectured in \cite{Bernard:1990cw,Zamolodchikov:1991vh,Fendley:1993xa} based on bootstrap, i.e.~by imposing unitarity, crossing symmetry and Yang-Baxter equation. They take the Restricted-Solid-On-Solid (RSOS) form \cite{Andrews:1984af} and read 
\begin{equation}
\resizebox{\hsize}{!}{$
    \widehat{S}^{ab}_{dc}(\theta)=Z(\theta)\(\frac{\d_a \d_c}{\d_b \d_d}\)^{\frac{i\theta}{2\pi}}\[ \sqrt{\dfrac{\d_a \d_c}{\d_b \d_d}}\, \sinh\(\frac{\theta}{n}\)\, \delta_{bd} + \sinh\(\frac{i\pi-\theta}{n}\)\, \delta_{ac}\, \], 
    $}\label{eq:Smatrixwrong}
\end{equation}
    Here $a,b,c,d=0,1/2,\ldots,n/2-1$ label the $n-1$ vacua (adjacent vacua differ by 1/2) and $\d_a=\frac{\sin \[(2a+1)\pi/n\]}{\sin (\pi/n)}$. The prefactor $Z(\theta)$ is crossing symmetric and is necessary for unitarity (see Supplemental Material). As we see below, these S-matrices turn out to be incompatible with non-invertible symmetries preserved along the flow.
 \subsubsection{Ward identity}
When the theory has a line $\mathcal{L}$ as its symmetry, the S-matrix satisfies the Ward identity (WI),
\begin{equation}
\sum\limits_g\; \includegraphics[height=2.3cm,valign=c]{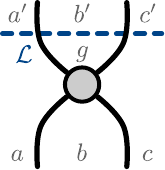} = \sum\limits_g\; \includegraphics[height=2.3cm,valign=c]{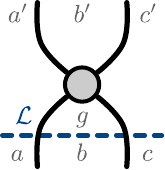}\label{eq:STDL}
\end{equation}
WI for non-invertible symmetries relates a single matrix element to a sum of matrix elements and imposes highly non-trivial constraints on the amplitudes\footnote{See Supplemental Material for explicit forms of constraints for the flow from tricritical Ising model.}. 

As we will derive in section \ref{subsec:symmetryaction}, the action of the symmetry line on a kink is given by (cf. \cite{Aasen:2020jwb})
\begin{equation}\label{eq:symaction}
\includegraphics[height=2.2cm,valign=c]{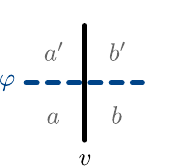}=\(\d_a\d_{a'}\d_{b}\d_{b'}\)^{1/4} 
    \begin{bmatrix}
     \varphi & a' & a\\
    v & b & b'
    \end{bmatrix}\,.
\end{equation}
 Using \eqref{eq:symaction}, we found that the S-matrices in the literature \eqref{eq:Smatrixwrong} {\it fail} to satisfy \eqref{eq:STDL}. This poses a sharp puzzle; mutual incompatibility of unitarity, crossing symmetry, Yang-Baxter equation and non-invertible symmetry.

\subsubsection{A new proposal}
Since integrability and non-invertible symmetry are established facts in conformal perturbation theory \cite{Zamolodchikov:1989hfa,Chang:2018iay}, and unitarity is fundamental, the only viable option is to relinquish crossing symmetry. Indeed we found S-matrices that satisfy all but crossing symmetry:  
\begin{equation}\label{eq:SRSOS}
\resizebox{\hsize}{!}{$
    S^{ab}_{dc}(\theta)=Z(\theta)\[ \sqrt{\dfrac{\d_a \d_c}{\d_b \d_d}}\, \sinh\(\frac{\theta}{n}\)\, \delta_{bd} + \sinh\(\frac{i\pi-\theta}{n}\)\, \delta_{ac}\, \], 
    $}
\end{equation}
These differ from \eqref{eq:Smatrixwrong} by a factor\footnote{The factor drops out in the derivation of Thermodynamic Bethe ansatz (TBA), so that both amplitudes are compatible with known TBA results.} 
$\( \frac{\d_a \d_c}{\d_b \d_d} \)^{\frac{i\theta}{2\pi}}$ and obey the modified crossing relation,
\begin{equation}\label{eq:modifiedcrossing}
    S^{ab}_{dc}(\theta)=\sqrt{\dfrac{\d_a \d_c}{\d_b \d_d}} \, S^{bc}_{ad}(i\pi-\theta)\,.
\end{equation}
We propose that \eqref{eq:SRSOS} is the correct S-matrix of this theory and provide a physical explanation of the modified crossing \eqref{eq:modifiedcrossing} in section \ref{subsec:modified}.

\subsubsection{\texorpdfstring{$\phi_{2,1}$-deformation of tricritical Ising model}{phi21-deformation of tricritical Ising model}}
Another example where  standard crossing is incompatible with non-invertible symmetries is the  $\phi_{2,1}$-deformation of the tricritical Ising model. In this case there are two vacua \cite{Lassig:1990xy} and the flow preserves the Fibonacci fusion category with elements $\{1,\mathcal W\}$ \cite{Chang:2018iay}. The spectrum consists of a kink, an anti-kink and a bound state, all with equal mass due to the non-invertible symmetry \cite{Cordova:2024vsq}. The S-matrix compatible with non-invertible symmetries is
 \begin{equation}
 \resizebox{\hsize}{!}{$
     S^{ab}_{dc}(\theta) = R(\theta) \[ \sqrt{\frac{\d_a \d_c}{\d_b \d_b}} \sinh\(\frac{\theta}{5/9}\) \delta_{bd} +  \sinh\(\frac{i\pi-\theta}{5/9}\)\delta_{ac} \] \,,
     $}
 \end{equation}
where $a,b,c,d=\{1,\mathcal{W}\}$ and $d_{1}=1$ and $d_{\mathcal{W}}=\frac{1+\sqrt{5}}{2}$. The overall factor $R(\theta)$ is given in the Supplemental Material. The amplitude obeys the same modified crossing \eqref{eq:modifiedcrossing}. 

In the literature, several different proposals exist for the S-matrix of this theory \cite{Zamolodchikov:1990xc,Smirnov:1991uw,Colomo:1991gw,Klassen:1992qy}. Our result supports the one in \cite{Smirnov:1991uw,Colomo:1991gw}, whose validity has not been established in the past as it violates standard crossing.

\subsection{\texorpdfstring{Perturbed $SU(2)_1$ WZW model}{Perturbed SU(2)1 WZW model} \label{subsec:SU2WZW}}
We next consider the $J \bar{J}$-deformed $SU(2)_1$ WZW model \cite{Ahn:1990gn,bernard1991quantum}; a prime example where standard crossing clashes with the $\mathbb{Z}_2$ 't Hooft anomaly. 
 The undeformed theory has $G=\frac{SU(2)_L \times SU(2)_R}{\bZ_2}$ symmetry, which reduces to $ \bZ_2\times SO(3)_V$ upon deformation,\footnote{We thank Shu-Heng Shao and Yifan Wang for noting an error in the previous version of this section.} with the $\bZ_2$ symmetry being anomalous. Due to the anomaly, the model cannot be trivially gapped, and has two degenerate vacua \cite{Ahn:1990gn}.
Alternatively it can be described as a compact scalar $X \sim X + 2\pi R$ at the self-dual radius $R=\sqrt{2}$ with the deformation,
\be \label{eq:cos2def}
J^a \bar{J}_a = \cos\left(2 X/R \right) \, ,
\ee
which has two minima at $X= \pm \pi R/2$.
This breaks the $U(1)_m \times U(1)_w$ symmetry to $(\bZ_2)_m \times U(1)_w$ with the anomalous $\bZ_2$ being the diagonal $\bZ_2^{\text{diag}}$. This description allows one to write kink-creation operators near the UV fixed point\footnote{These operators are singled out by imposing 1.~they have a half-unit of winding since kinks interpolate between two vacua, 2.~they need to be purely holomorphic or anti-holomorphic since kinks are massless in the UV.} \cite{Klassen:1992eq}
\be
\resizebox{0.85\hsize}{!}{$
K_\pm = \left\{ V_{\frac{1}{2}, \frac{1}{2}} \, , V_{\frac{1}{2}, - \frac{1}{2}} \right\} \, , \quad
\bar{K}_\pm = \left\{ V_{-\frac{1}{2}, -\frac{1}{2}} \, , V_{-\frac{1}{2}, \frac{1}{2}} \right\} \, ,$}
\ee
where $V_{m,w}$ is a vertex operator with momentum $m$ and winding $w$. As they have half-units of $m$ and $w$, these operators are endpoints of $\mathbb{Z}_2^{\rm diag}$, indicating that kink worldlines act as symmetry lines of $\mathbb{Z}_2^{\rm diag}$.

\begin{figure}
    \centering
    \includegraphics[width=.8\linewidth]{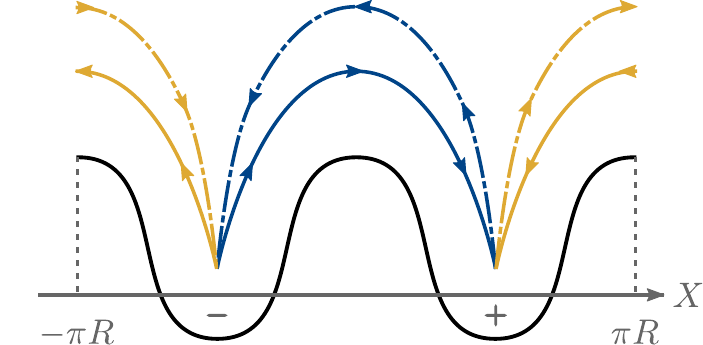}
    \caption{Kinks (solid lines) and anti-kinks (dashed lines) connecting the minima of the potential. Both come with two-fold degeneracy (blue and yellow) due to the Kramers degeneracy.}
    \label{fig:kinksWZW}
\end{figure}

As shown in \cite{Hason:2020yqf, Cordova:2019wpi}, the $\mathbb{Z}_2$ anomaly induces the time reversal anomaly $T^2=-1$ on such kink worldlines, implying the two-fold Kramers degeneracy of kinks. Here it simply corresponds to two different ways of connecting the two vacua, see Figure \ref{fig:kinksWZW}.

The S-matrix of these (anti-)kinks has a tensor-product structure \cite{Ahn:1990gn,bernard1991quantum}, $S_{{\rm RSOS}_2}\otimes \check{S}$,
where $S_{{\rm RSOS}_2}$ is a trivial factor that distinguishes the two vacua (denoted by $\pm$)
\be
S_{{\rm RSOS}_2}: \,\includegraphics[height=1.2cm,valign=c]{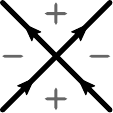}\,=\,\includegraphics[height=1.2cm,valign=c]{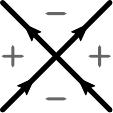}\,=1\,, 
\ee
while $\check{S}$ is the scattering among Kramers pairs \cite{Ahn:1990gn,bernard1991quantum,Vu:2019qxt}

\be
\check{S}^{ab}_{dc}(\theta)=\includegraphics[height=1.7cm,valign=c]{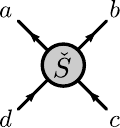}=S_0(\theta)\,\frac{\theta\, \delta_{c}^{a}\delta_{d}^{b}-i\pi\, \delta_{c}^{b}\delta_{d}^{a}}{\theta-i\pi }\,,
\ee
with $a,b,c,d=\{1,2\}$ and
\be
S_0(\theta)=-\frac{\Gamma(1-\theta/2\pi i)\Gamma(1/2+\theta/2\pi i)}{\Gamma(1+\theta/2\pi i)\Gamma(1/2-\theta/2\pi i)}\,.
\ee

One can verify that $\check{S}$ violates standard crossing rule by a sign:
\be\label{eq:modifiedcrossZ2}
\check{S}^{ab}_{dc}(\theta)=-C_{cc^{\prime}}\check{S}^{bc^{\prime}}_{a^{\prime}d}(i\pi -\theta)C^{a^{\prime }a}\,,
\ee
where $C$ is the charge conjugation matrix satisfying $C^{T}=-C$ and $C_{12}=-C_{21}=-1$. Note that $C^2=-\mathbb{1}$ as expected from the time-reversal anomaly on the worldline.\footnote{This follows from $CT$ being anomaly-free on the worldline.} This violation of crossing was already noted in the literature \cite{bernard1991quantum} but a physical origin was not elucidated. In section \ref{subsec:modified}, we show that the extra sign in \eqref{eq:modifiedcrossZ2} is due to the $\mathbb{Z}_2$ anomaly.

A couple of comments are in order: 1.~the deformation \eqref{eq:cos2def} is different from the standard sine-Gordon model with the potential $\cos (X/R)$. In the latter case, the vacuum is unique and there is no $\mathbb{Z}_2$ anomaly. Hence it satisfies standard crossing. 2.~the compact boson with the deformation \eqref{eq:cos2def} (called SG$(\beta,2)$ in \cite{Klassen:1992eq,Klassen:1992qy}) is $\mathbb{Z}_2$-anomalous and obeys the modified crossing, even at $R\neq \sqrt{2}$ . See the Supplemental Material for details.

\section{Derivation of symmetry action and modified crossing}\label{sec:modified}
\subsection{Action of symmetry lines}\label{subsec:symmetryaction}
    We now derive the action of symmetry lines on kinks \eqref{eq:symaction} by studying states on a line with specified boundary conditions at infinity indicating the vacua\footnote{This is essentially the same as `open' Hilbert space discussed e.g.~in \cite{Huang:2021zvu}.}. We denote them by $|\psi\rrangle$ to distinguish them from states on a circle.
    
    We begin by evaluating a path integral on a large Euclidean disk with radius $R(\gg m^{-1})$, where fields approach the vacuum `0' at the boundary. This computes the norm of $|0\rrangle$ on a line, which we normalize to be $1$. 
    \be
     \llangle 0|0\rrangle=\includegraphics[height=1.2cm,valign=c]{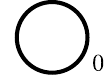}=1\,.
    \ee
    Since the vacua transform as the regular representation (cf.~\eqref{eq:genvac}), the norms of other vacua $|a\rrangle$ can be computed by inserting a loop of $\mathcal{L}_a$ on the disk. 
    Since $\langle \mathcal{L}_a\rangle=\d_a$, we get\footnote{In CFTs, $\d_a$ is related to a ratio of $g$-functions $\d_a = g_a/g_0$.}
    \be|a\rrangle=\includegraphics[height=.75cm,valign=c]{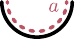}\,,\quad \llangle a|a\rrangle=\includegraphics[height=1.5cm,valign=c]{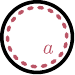}=\d_a\,.
    \ee

    Next consider a matrix element of $\mathcal{L}_{\varphi}$ on normalized states, $\frac{\llangle b|\mathcal{L}_{\varphi}|a\rrangle}{\sqrt{\llangle a|a\rrangle\llangle b|b\rrangle}}$. This can be represented in the path integral as
    \be
    \frac{\llangle b|\mathcal{L}_{\varphi}|a\rrangle}{\sqrt{\llangle a|a\rrangle\llangle b|b\rrangle}} = \frac{1}{\sqrt{\d_a\d_b}} \,\includegraphics[height=1.6cm,valign=c]{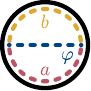}
    \ee
    Using \eqref{eq:Lortho} and \eqref{eq:Lqd}, we evaluate the network of lines as
     \be
      \label{eq:threenetwork}\,\includegraphics[height=1.6cm,valign=c]{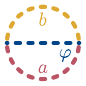} =\sqrt{\d_a\d_b\d_{\varphi}}\, N_{\varphi\, a}{}^{b}\,.
     \ee
    This shows that, when computing the action on an infinite line, each $\mathcal{L}_{\varphi}$ needs to be divided by $\sqrt{\d_{\varphi}}$ in order to realize the correct action: 
    \be
    \hat{\mathcal{L}}_{\varphi}= \frac{1}{\sqrt{\d_\varphi}} \; \includegraphics[width=1.5cm,valign=c]{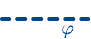}
    \ee

    Finally we consider a state with a single kink between two vacua, $a$ and $b$. The kink interpolates between neighboring vacua and, in the IR limit, its effect on the vacua is identical to that of the symmetry line $\mathcal{L}_{v}$,  with $v=\frac{1}{2}$ for the $\phi_{1,3}$-deformations and $v=\mathcal{W}$ for the $\phi_{2,1}$-deformation of the tricritical Ising model.\footnote{In the generic case, identifying the line 
$v$ involves determining the UV CFT counterpart of the kink creation operator, which is an operator in the 
$v$-twisted sector.} Thus, the state can be represented by the path integral:
    \be
    |a;b\rrangle=  \; \includegraphics[height=.8cm,valign=c]{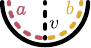} 
    \ee
    Its norm can be computed using \eqref{eq:threenetwork} as $\llangle a ;b|a;b\rrangle=\sqrt{\d_a\d_b\d_v}$. A matrix element of $\hat{\mathcal{L}}_{\varphi}$ on normalized states can be computed using \eqref{eq:Lortho}, \eqref{eq:Lcomp} and \eqref{eq:Fsymb} as\footnote{Note that the lines $v$ and $\varphi$ do not intersect (i.e.~there is no junction between them). This choice  is motivated by 1.~the symmetry line should not act directly on kinks; rather its action on kinks should derive automatically from the action on vacua. 2.~Eq.~\eqref{eq:crossaction} coincides with the action of symmetry lines on lattice models studied in \cite{Aasen:2020jwb}.}
    
    \be\label{eq:crossaction}
    \begin{aligned}
     \resizebox{0.4\hsize}{!}{$
     \frac{\llangle a^{\prime};b^{\prime}|\hat{\mathcal{L}}_{\varphi}|a;b\rrangle}{\sqrt{\llangle a;b|a;b\rrangle\llangle a^{\prime};b^{\prime}|a^{\prime};b^{\prime}\rrangle}}
     $}&=
     \resizebox{0.54\hsize}{!}{$
     \(\d_a\d_{a'}\d_{b}\d_{b'}\d_v^2 \d_\phi^2\)^{-\frac{1}{4}}\;  \includegraphics[height=2.2cm,valign=c]{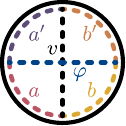}
     $}\\
     &= \(\d_a\d_{a'}\d_{b}\d_{b'}\)^{1/4} 
    \begin{bmatrix}
     \varphi & a' & a\\
    v & b & b'
    \end{bmatrix}\,,
     \end{aligned}
    \ee
    which gives \eqref{eq:symaction}. 

\subsection{Modified crossing}\label{subsec:modified}
We now present a general derivation of modified crossing rules of S-matrices. As we see below, the main reasons for the modification are
\begin{itemize}
\item The IR dynamics is described by a nontrivial TQFT.
\item Normalizations of in- and out-states receive corrections from the TQFT degrees of freedom.
\item The corrections are different between $s$- and $t$-channels.
\end{itemize}
The derivation does not rely on integrability. Thus the modified crossing rule \eqref{eq:final} should be applicable to non-integrable theories as well.
\subsubsection{Derivation}
The Lehmann–Symanzik–Zimmermann (LSZ) reduction formula, commonly used to derive S-matrices from local operator correlation functions, is not applicable in our cases due to the non-locality of kink-creation operators attached to lines. Instead, we adopt an alternative formalism linking the S-matrix to correlation functions on two Cauchy slices \cite[Ch.~5]{Itzykson:1980rh}. This is expressed as \cite{Dubovsky:2017cnj}\footnote{Strictly speaking, the formula has not been derived for non-local operators such as kink-creation operators. However it was proven useful in another context in which non-locality is present \cite{Dubovsky:2017cnj}, and we assume that it also holds in our case. Also, here we are assuming implicitly that operator insertions in $G(\{x\})$ are approriately normalized so that it gives unitary S-matrix.}:
\be
\resizebox{\hsize}{!}{$
S(\{p_i\})=\lim_{T\to\infty}\int \prod_{j}dv_je^{ip_jx(v_j)}\prod_{k}(n_k\cdot \overset{\leftrightarrow}{\partial}_{x_k})G(\{x(v_k)\})\,,
$}
\ee
where $x^{\mu}(v)$ parametrizes past and future Cauchy surfaces, $T$ denotes the time separation between them, $n_k$ are unit normal vectors at points $v_k$, and $G(\{x(v_k)\})$ represents a correlation function with kink-creation operators inserted at $v_k$. 

By analytic continuation, $G(\{x(v_k)\})$ is related to a correlation function on a large Euclidean disk with appropriate boundary condition changing operators that create kinks. Thus the S-matrix of kinks can be expressed schematically as
\be
S^{ab}_{dc}(\theta)\propto\; \includegraphics[height=1.5cm,valign=c]{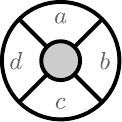}\; \Bigg\lvert _\text{analyt. cont.}
\ee
However, to ensure standard unitarity, we must take into account normalizations of in- and out-states,
\be
|\text{in}\rrangle=\includegraphics[height=1.1cm,valign=c]{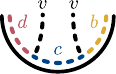},\quad |\text{out}\rrangle=\includegraphics[height=1.1cm,valign=c]{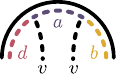}\,,
\ee
which are given by networks of lines, \footnote{Here we are stripping out the usual momentum dependent normalization of the two-particle state in terms of the (squared) center of mass energy $s=(p_1+p_2)^2$ $$\langle p_1',p_2'|p_1,p_2\rangle=(2\pi)^2\, 2\sqrt s \sqrt{s-4m^2}\,\delta^2(p_1+p_2-p_1'-p_2')\,.$$} 
\be
\begin{aligned}
 \llangle \text{in}|\text{in}\rrangle=\includegraphics[height=1.4cm,valign=c]{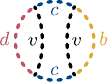}\,,\quad 
 \llangle \text{out}|\text{out}\rrangle=\includegraphics[height=1.4cm,valign=c]{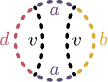}\,.
 \end{aligned}
\ee
Physically, these represent corrections to normalizations of in- and out-states due to the TQFT dynamics.
Thus the S-matrix between normalized states, which satisfies standard unitarity, is given by
\be
S^{ab}_{dc}(\theta)=\frac{\quad\quad\quad\includegraphics[height=1.3cm,valign=c]{figures/Sdisc.pdf}\; \Bigg\lvert _\text{analyt. cont.}}{\sqrt{\includegraphics[height=1.2cm,valign=c]{figures/innorm.pdf}\;\,\includegraphics[height=1.2cm,valign=c]{figures/outnorm.pdf}}}\,.
\ee
Notably, while the numerator is crossing symmetric, the denominator changes with the channels one considers. This leads to the modification of crossing symmetry
\be\label{eq:final}
S^{ab}_{dc} (\theta)=\sqrt{\frac{\includegraphics[height=1.2cm,valign=c]{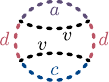}\;\,\includegraphics[height=1.2cm,valign=c]{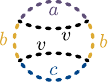}}{\includegraphics[height=1.2cm,valign=c]{figures/innorm.pdf}\;\,\includegraphics[height=1.2cm,valign=c]{figures/outnorm.pdf}}}\; S^{bc}_{ad}(i\pi-\theta)\,.
\ee
For the gapped flows from minimal models, the ratio of the networks in \eqref{eq:final} gives the modified crossing factor found in \eqref{eq:modifiedcrossing}. For the case with the $\mathbb{Z}_2$ anomaly, the ratio  gives the $F$-symbol $\epsilon=-1$ (see \eqref{eq:Z2Fsymb}), reproducing the sign in the crossing relation \eqref{eq:modifiedcrossZ2}.

\section{Conclusion}
We initiated the study of the interplay among non-invertible symmetries, anomalies, and S-matrices, focusing on integrable flows in $1+1$ dimensions. Our findings show that the crossing symmetry of S-matrices is modified when the IR phase is governed by a nontrivial TQFT, as the TQFT affects the normalization of in- and out-states.

There are numerous future directions to explore:
\begin{itemize}
\item Testing our proposed S-matrices by computing physical observables that distinguish ours from those in the literature: A prime candidate is the finite-volume spectrum with an insertion of a non-invertible defect (i.e.~defect Hilbert space) \cite{inprogress}.
\item In our $\bZ_2$ example the modification of crossing symmetry was a consequence of the 't Hooft anomaly. 
In cases with non-invertible symmetry, the 't Hooft anomaly prevents the flow to a trivially gapped phase while the modification of crossing symmetry is captured by a ratio of quantum dimensions, which are relative Euler terms \cite{Bhardwaj:2023idu}. It would be interesting to clarify the precise relationship between 't Hooft anomalies and the modified crossing.
\item Finding examples of modified crossing in higher dimensions: In 2+1 dimensions, modifications have been observed in Chern-Simons matter theories \cite{Mehta:2022lgq}, which exhibit nontrivial TQFT dynamics in the IR. It is important to understand these findings in a unified (categorical) framework alongside our case. Additionally, our analysis suggests that such modifications can occur even in theories where TQFT degrees of freedom are not apparent in the UV Lagrangian description. Identifying such examples in 2+1 dimensions and beyond would be of great interest. Furthermore examples of RG flows preserving non-invertible symmetries in $3+1$ dimensions have been constructed \cite{Kaidi:2021xfk,Choi:2021kmx,Choi:2022zal,Choi:2022jqy,Cordova:2022ieu,Choi:2022rfe,Damia:2023ses} . It would be interesting to see the consequence of non-invertible symmetries on S-matrices in such examples.
\item Performing the S-matrix bootstrap analysis  \cite{Paulos:2016but,Cordova:2018uop,He:2018uxa} of theories with non-invertible symmetries \cite{inprogress2}: Particularly interesting are theories with the Haagerup fusion category \cite{haagerup1994principal,asaeda1999exotic,
Huang:2020lox}, for which the QFT realization is not fully understood. See Supplemental Material for crossing and unitarity in the projector basis, which will be useful for performing such analyses. Additionally, bounding the UV central charge of such theories via form factor bootstrap  \cite{Karateev:2019ymz,Cordova:2023wjp} could be valuable, especially given indications that the fixed point has a central charge 2 in certain lattice realizations \cite{Vanhove:2021zop,Huang:2021nvb}. 
\item We focused on examples in which the vacua are in the regular representation. Extending to other representations should be straightforward by combining our analysis with that in \cite{Huang:2021zvu}. We expect that the natural generalization to IR vacua in arbitrary representations is  given by a factor 
\be
\epsilon_v \, \sqrt{\frac{\tilde{d}_a \tilde{d}_c}{\tilde{d}_b \tilde{d}_d}} \, ,
\ee
where $\epsilon_v=\pm 1$ is the Frobenius-Schur indicator of $v$ (i.e.~its time reversal anomaly) and $\tilde{d}_a$ is the disk partition function \cite{Huang:2021zvu}.

\item It was argued that the scattering of monopoles and fermions in $3+1$ dimensions violates crossing symmetry \cite{Csaki:2020inw}, alongside indications of significant roles played by non-invertible symmetries \cite{vanBeest:2023dbu,vanBeest:2023mbs}. Establishing a connection with our analysis would be interesting.
\item The nontrivial TQFT dynamics in our examples serve as toy models for the soft IR dynamics of photons and gravitons \cite{Strominger:2017zoo}. Just as in our examples, the Faddeev-Kulish dressing \cite{Kulish:1970ut} modifies the norms of in- and out-states. Furthermore the IR phases of electrodynamics and gravity possess exotic (generalized) symmetries \cite{Hofman:2018lfz,Benedetti:2021lxj,Hinterbichler:2022agn}. Can the categorical language offer a suitable framework for understanding these soft dynamics?
\end{itemize}

\subsection*{Acknowledgement}
We thank Lorenzo Di Pietro, Thomas Dumitrescu, Hofie Hannesdottir, Po-Shen Hsin, Ying-Hsuan Lin, Jiaxin Qiao, Giovanni Rizi, Didina Serban, Shu-Heng Shao, Roberto Tateo and Yifan Wang for helpful discussions.
CC is
supported by STFC grant ST/X000761/1. This research was supported in part by grant NSF PHY-2309135 to the Kavli Institute for Theoretical Physics (KITP).
\bibliographystyle{JHEP}
\bibliography{intebib}
\section*{Supplemental Material}
\begin{appendices}
\subsection{\texorpdfstring{$F$-symbols}{F-symbols}}\label{app:Fsymbols}
As \eqref{eq:Ftetra} shows, F-symbols and tetrahedral symbols are proportional to each other. 
For the $\mathcal{A}_{n-1}$ category discussed in the main text the tetrahedral symbols are the same as $q$-deformed Wigner 6j-symbols up to a sign:
\begin{gather}
    \begin{bmatrix}
     a & b & c\\
    d & e & f
    \end{bmatrix}= 
    (-1)^p \begin{Bmatrix}
     a & b & c\\
    d & e & f
    \end{Bmatrix}_q \,,\quad q=e^{2\pi i/n}\,,\\
    \resizebox{\hsize}{!}{$
    p=\frac{1}{2}\[3 (a + b + c + d + e + f)^2-(a+d)^2-(b+e)^2-(c+f)^2\]\,. \nonumber
    $}
\end{gather}
The $q$-deformed Wigner 6j-symbols are given in terms of the quantum dimensions $[a]=[a]_q=\d_a=\frac{\sin \[(2a+1)\pi/n\]}{\sin (\pi/n)}$ as follows
\begin{equation}
    \begin{split}
    &\begin{Bmatrix}
     a & b & c\\
    d & e & f
    \end{Bmatrix}_q = \Delta(a, b, c)\, \Delta(a, e, f)\, \Delta(d, b, f)\, \Delta(d, e, c) \times
    \\
    &\resizebox{.8\hsize}{!}{$\sum\limits_z
    \frac{(-1)^z\[z+1\]}{\[a+b+d+e-z\]\[a+c+d+f-z\]\[b+c+e+f-z\]}\times
    $}\\
    &\resizebox{.8\hsize}{!}{$
    \frac{1}{\[-a-b-c+z\]\[-c-d-e+z\]\[-b-d-f+z\]\[-a-e-f+z\]}
    $}\,,
    \end{split} 
\end{equation}
where the sum is over half integers $z$ such that the summand is free of divergences and  

\begin{equation}
  \Delta(a,b,c)=
  \begin{cases}
      \sqrt{\frac{[a + b - c] [a - b + c] [-a + b + c]}{[1 + a + b + c]}} & \text{if}\ N_{abc}=1 \\
      0 & \text{otherwise}\,.
      \end{cases}
\end{equation}

\subsection{Scalar factors of exact S-matrices}\label{ap:def}
The prefactor in \eqref{eq:SRSOS} is given by \cite{Fendley:1993xa}
\begin{equation}\label{eq:Zdef}
\resizebox{\hsize}{!}{$
    Z(\theta)=\frac{1}{\sinh{\frac{\theta-i\pi}{n}}} \exp\left\{\frac{i}{2} \int\limits_{-\infty}^\infty \frac{dk}{k}\, \sin{k\theta}\, \frac{\sinh{\frac{k\pi}{2}(n-1)}}{\sinh{\frac{n k\pi}{2}\, \cosh{\frac{k\pi}{2}}}} \right\}\,.
    $}
\end{equation}
It is crossing symmetric $Z(\theta)=Z(i\pi-\theta)$ and satisfies $Z(\theta)Z(-\theta)=\left(\sinh{\frac{\theta-i\pi}{n}}\sinh{\frac{-\theta-i\pi}{n}}\right)^{-1}$.
Similarly for the $\phi_{2,1}$ deformation the prefactor $R(\theta)$ is crossing symmetric and ensures unitarity \cite{Smirnov:1991uw,Klassen:1992qy} 
\begin{gather}
\resizebox{\hsize}{!}{$
    R(\theta)=\dfrac{-1}{\sinh\dfrac{\theta+i\pi}{5/9}} f_{-2/5}\(9\theta/5\) f_{3/5}(9\theta/5) F_{-1/9}(\theta) F_{2/9}(\theta) \,, 
    $}\label{eq:Rdef} \\
     R(\theta) R(-\theta) = \left(\sinh{\frac{\theta-i\pi}{5/9}}\sinh{\frac{\theta+i\pi}{5/9}}\right)^{-1} \,.
\end{gather}
The building blocks are $f_\alpha(\theta)=\frac{\sinh\(\frac{\theta+i\alpha\pi}{2}\)}{\sinh\(\frac{\theta-i\alpha\pi}{2}\)}$ and $F_\alpha$ are the Castillejo-Dalitz-Dyson (CDD) factors $F_\alpha(\theta)=-f_\alpha(\theta)f_\alpha(i\pi-\theta)$.
\subsection{\texorpdfstring{Non-invertible symmetries of $\phi_{r,s}$ deformations}{Non-invertible symmetries of phirs deformations}}
Here we study symmetry lines preserved by the $\phi_{r,s}$ deformation of the $n$-th minimal model $\mathcal{M}_n=\mathcal M_{n+1,n}$:
\be
 \cM_{n} \longrightarrow \, \cM_{n} + \lambda \int \phi_{r, \, s} \, .
\ee
First, recall that the action of Verlinde lines $\mathcal L_{r', \, s'}$ onto a primary $\phi_{r, \, s}$ is given by (see e.g. \cite{Chang:2018iay} for a review):
\begin{equation}\label{eq:LactionPhi}
    \includegraphics[height=1.4cm,valign=c]{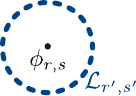}=\frac{\mathcal{S}_{r',  s' ;\,r ,  s}}{\mathcal{S}_{1,  1 ; \, r,  s}}\includegraphics[height=1.4cm,valign=c]{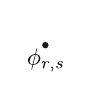}
\end{equation}
where the modular $S$-matrix is given by
\be
\resizebox{\hsize}{!}{$
\mathcal S_{r',  s';\, r , s} =  \sqrt{\frac{8}{n(n+1)}} \, (-)^{1 + r s' + r' s} \sin \left(  \frac{\pi n r r'}{n+1} \right) \sin\left(  \frac{\pi(n+1)s s'}{n}  \right) \,. $}
\ee
The line $\mathcal L_{r', \, s'}$ is preserved along the flow if and only if its action on the deforming operator is trivial. That is, if 
\begin{equation}
    \eqref{eq:LactionPhi}\,
    =\;\includegraphics[height=1.4cm,valign=c]{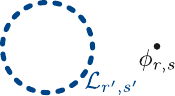}\,
    =\frac{\mathcal{S}_{r',  s' ;\,1 ,  1}}{\mathcal{S}_{1,  1 ; \, 1,  1}}\includegraphics[height=1.4cm,valign=c]{figures/Phi0.pdf}
\end{equation}

In the case of the $\phi_{1,3}$ discussed in the main text, the above equation reduces to the condition: 
\be
\cos\left( \frac{2 \pi }{n} s'\right) = \cos\left( \frac{2 \pi}{n} \right) \, , 
\ee
whose only solution (up to the $\bZ_2$ action on the Kac table) is $s'=1$. Thus the $\phi_{1, \, 3}$ deformation preserves exactly the $\cA_{n-1}$ subalgebra given by the $\mathcal L_{r',1}$ lines. Following the flow $\cM_{n}\rightarrow \cM_{n-1}$ this is mapped into the $\tilde{\cA}_{n-1}$ algebra in the IR minimal model, generated by $\cL_{1,2}$.
\subsection{Examples of Ward identities}
Here we show explicitly some of the Ward identities \eqref{eq:STDL} the S-matrix for the $\phi_{1,3}$ deformation of the $n$-th minimal model $\mathcal M_n$ should satisfy. In this gapped flow, the $n-1$ vacua are labelled by $a=0,\frac{1}{2},\ldots,\frac{n-2}{2}$. The only non trivial symmetry line present for any $n$ is the $\mathbb{Z}_2$ (invertible) defect $\eta$. It maps the first vacuum to last one and more generally $a\leftrightarrow \frac{n-2}{2}-a$. Since the previous map is one to one, the Ward identities relate two S-matrix elements, for instance:
\begin{equation}
   \eta:\quad S^{l-\frac{1}{2},\, l}_{l,\, l+\frac{1}{2}}(\theta)=S^{\frac{n}{2}-l-\frac{1}{2},\,\frac{n}{2}- l-1}_{\frac{n}{2}-l-1,\,\frac{n}{2}- l-\frac{3}{2}}(\theta)\,.
\end{equation}
These Ward identities are satisfied both for the amplitude previously written in the literature \eqref{eq:Smatrixwrong} and our proposal \eqref{eq:SRSOS}. It follows from the fact that the quantum dimensions are the same between the vacua $\d_a=\d_{\frac{n-2}{2}-a}$.
Now let us check a WI for the simplest non-invertible defect $\mathcal N$ present in the flow from tricritical Ising $\mathcal M_4$. The action of the symmetry lines on the three vacua $a=0,\frac{1}{2},1$ can be represented in matrix form as follows
\begin{equation}
    \mathbb 1=\begin{pmatrix}
1 & 0 & 0\\
0 & 1 & 0\\
0 & 0 & 1
\end{pmatrix}\,,\quad
\mathcal N=\begin{pmatrix}
0 & 1 & 0\\
1 & 0 & 1\\
0 & 1 & 0
\end{pmatrix}\,,\quad
\eta=\begin{pmatrix}
0 & 0 & 1\\
0 & 1 & 0\\
0 & 0 & 1
\end{pmatrix}\,.
\end{equation}
One of the WI for the $\mathcal N$ line is
\begin{equation*}\includegraphics[height=2.2cm,valign=c]{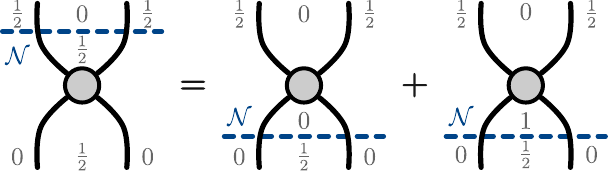}\end{equation*}
\begin{equation}
  \mathcal N:\quad  S^{\frac{1}{2} 0}_{0 \frac{1}{2}}(\theta) = S^{0 \frac{1}{2}}_{\frac{1}{2} 0}(\theta) +S^{0 \frac{1}{2}}_{\frac{1}{2} 1}(\theta)\,,
\end{equation}
which our proposal \eqref{eq:SRSOS} satisfies:
\begin{equation}
\resizebox{\hsize}{!}{$
 \sqrt{2}\sinh\frac{\theta}{4}+\sinh\frac{i\pi-\theta}{4}=\frac{1}{\sqrt{2}}\sinh\frac{\theta}{4}+\sinh\frac{i\pi-\theta}{4}+\frac{1}{\sqrt{2}}\sinh\frac{\theta}{4}\,,
 $}
\end{equation}
but the amplitude \eqref{eq:Smatrixwrong} does not:
\begin{equation}
\begin{split}
 &\(\sqrt{2}\sinh\frac{\theta}{4}+\sinh\frac{i\pi-\theta}{4}\)\sqrt{2}^{\,i\theta/\pi}\\
 &\neq\(\frac{1}{\sqrt{2}}\sinh\frac{\theta}{4}+\sinh\frac{i\pi-\theta}{4}+\frac{1}{\sqrt{2}}\sinh\frac{\theta}{4}\)\(\frac{1}{\sqrt{2}}\)^{\,i\theta/\pi}\,.
 \end{split}
\end{equation}
In a similar fashion, the WI for non-invertible defects for any $n$ are spoiled by the $\(\frac{\d_a\d_c}{\d_b\d_d}\)^{\frac{i\theta}{2\pi}}$ factors in \eqref{eq:Smatrixwrong}.
\subsection{Crossing and unitarity in the projector basis}
We can construct a basis of projectors of two $v=1/2$ lines into a fusion channel $\chi$ satisfying
\begin{gather}
    P_\chi P_{\chi'} = \delta_{\chi \chi'} P_\chi \,,\label{eq:Pproj}\\
    \sum_\chi P_\chi = \unit \,. \label{eq:Punit}
\end{gather}
These projectors can be written as\footnote{Here we are considering categories for which the fusion coefficients are $N_{ab}^c=0,1$.} 
\begin{equation}\label{eq:Pdef}
    (P_\chi )^{ab}_{dc}= \includegraphics[height=2cm,valign=c]{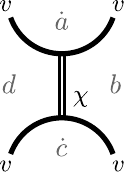}=\sqrt{\d_a \d_c}\, \d_\chi 
    \begin{bmatrix}
     v & v & \chi\\
    d & b & a
    \end{bmatrix}
    \begin{bmatrix}
    v & v & \chi\\
    d & b & c
    \end{bmatrix} \,,
\end{equation}
where we have used dotted indices and double line notation to emphasize the prefactor $\sqrt{\d_a \d_c}\, \d_\chi$. Using orthogonality of the tetrahedral symbols 
\begin{equation}\label{eq:ortho}
    \sum_\chi \d_\chi 
    \begin{bmatrix}
     v & v & \chi\\
    d & b & a
    \end{bmatrix}
    \begin{bmatrix}
    v & v & \chi\\
    d & b & c
    \end{bmatrix} = \delta_{ac} \frac{N^v_{ba}N^v_{da}}{\d_a}\,,
\end{equation}
one can verify that that \eqref{eq:Pdef} indeed satisfies \cref{eq:Pproj,eq:Punit}. Explicitly, we have

\begin{equation}\label{eq:Pprojcheck}
\begin{split}
   &\sum_e (P_\chi )^{ab}_{de} \, (P_{\chi'} )^{eb}_{dc}=\\
   &\resizebox{\hsize}{!}{$=\sum\limits_e \sqrt{\d_a \d_c}\, \d_e \d_\chi \d_{\chi'} 
   \begin{bmatrix}
     v & v & \chi\\
    d & b & a
    \end{bmatrix}
    \begin{bmatrix}
    v & v & \chi\\
    d & b & e
    \end{bmatrix} 
    \begin{bmatrix}
     v & v & \chi'\\
    d & b & e
    \end{bmatrix}
    \begin{bmatrix}
    v & v & \chi'\\
    d & b & c
    \end{bmatrix} $}\\
    &=\delta_{\chi\chi'}\sqrt{\d_a \d_c}\, \d_\chi
    \begin{bmatrix}
     v & v & \chi\\
    d & b & a
    \end{bmatrix} \begin{bmatrix}
    v & v & \chi\\
    d & b & c
    \end{bmatrix}
    =\delta_{\chi\chi'} \, (P_\chi )^{ab}_{dc}\,,
\end{split}    
\end{equation}

\begin{equation}
    \sum_\chi (P_\chi)^{ab}_{dc}=\sum_\chi \sqrt{\d_a \d_c}\, \d_\chi 
    \begin{bmatrix}
     v & v & \chi\\
    d & b & a
    \end{bmatrix}
    \begin{bmatrix}
    v & v & \chi\\
    d & b & c
    \end{bmatrix}
    = \delta_{ac}\,.
\end{equation}
Note that the square root prefactor in \eqref{eq:Pdef} is crucial in the above equations.

The projectors make the categorical symmetry manifest, as they commute with the action of the symmetry lines as defined in \eqref{eq:symaction}: 
\begin{equation}
\begin{split}
   &\resizebox{.96\hsize}{!}{$
   \includegraphics[height=2cm,valign=c]{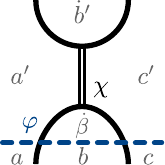} = \sum\limits_\beta \omega\sqrt{\d_\beta \d_b} 
    \begin{bmatrix}
     \varphi & a' & a\\
     v & b & \beta
    \end{bmatrix}
    \begin{bmatrix}
     \varphi & \beta & b\\
     v & c & c'
    \end{bmatrix}
    \(P_\chi\)^{b'c'}_{a'\beta}
    $}\\
    = & \resizebox{.96\hsize}{!}{$
     \sum\limits_\beta \omega \sqrt{\d_{b'} \d_\beta} 
    \begin{bmatrix}
     \varphi & a' & a\\
     v & \beta & b'
    \end{bmatrix}
    \begin{bmatrix}
     \varphi & b' & \beta\\
     v & c & c'
    \end{bmatrix}
    \(P_\chi\)^{\beta c}_{a b} = \includegraphics[height=2cm,valign=c]{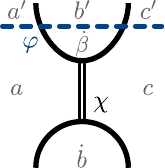} 
    $}
\end{split}
\label{eq:Pcommute}
\end{equation}
where $\omega=(\d_a\d_{a'}\d_c\d_{c'})^{1/4}$.
The previous equality follows from the pentagon identity satisfied by the tetrahedral symbols 
\begin{equation}\label{eq:pentagon}
\resizebox{\hsize}{!}{$
        \sum\limits_\beta \d_\beta 
        \begin{bmatrix}
         \varphi & a' & a\\
         y & b & \beta
        \end{bmatrix}
         \begin{bmatrix}
         \varphi & c' & c\\
         x & b & \beta
        \end{bmatrix}
        \begin{bmatrix}
         a' & c' & \chi\\
         x & y & \beta
        \end{bmatrix}= \begin{bmatrix}
         a & c & \chi\\
         x & y & b
        \end{bmatrix}
        \begin{bmatrix}
         a & c & \chi\\
         c' & a' & \varphi
        \end{bmatrix}\,.
        $}
\end{equation}

Having these projectors, we can write a two-particle amplitude compatible with the categorical symmetry as
\begin{equation}\label{eq:Sproj}
    S^{ab}_{dc}(\theta)=\sum_\chi A_\chi(\theta) (P_\chi)^{ab}_{dc}\,.
\end{equation}
Elastic unitarity of the S-matrix is satisfied if $A_\chi(\theta) A_\chi(-\theta)=1$.

In the expressions above we have chosen the s-channel decomposition. If we want to instead use the t-channel projectors we need to perform an F-move $\textit{and}$ multiply by $\sqrt{\frac{\d_a\d_c}{\d_b\d_d}}$. 
This factor is necessary to ensure the t-channel projectors obey \cref{eq:Pproj,eq:Punit}. The t-channel projectors are then given by
\begin{equation}
\resizebox{.96\hsize}{!}{$
    (P_\chi)^{bc}_{ad}= \includegraphics[width=2cm,valign=c]{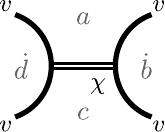}=\sqrt{\frac{\d_b\d_d}{\d_a\d_c}} \, \d_\chi \sum\limits_{\chi'}
    \begin{bmatrix}
     v & v & \chi\\
     v & v & \chi'
    \end{bmatrix}
    (P_{\chi'})^{ab}_{dc}\,.
    $}
\end{equation}

In the $A_{n-1}$ category discussed in the main text we can use the tetrahedral symbols defined in appendix~\ref{app:Fsymbols} to evaluate the two projectors $\chi=0,1$ \footnote{In verifying \eqref{eq:Pprojcheck} one encounters the sum over quantum dimensions $\sum_e\d_e$, however the values of $e$ are restricted to $b\pm1/2$. To write \eqref{eq:PRSOS} it is enough to use the identity $$\d_v d_b=d_{b-1/2}+d_{b+1/2}\,.$$}
\begin{equation}\label{eq:PRSOS}
    (P_0)^{ab}_{dc}=\frac{1}{\d_v} \sqrt{\frac{\d_a \d_c}{\d_b \d_d}}\delta_{bd}\,,\quad  (P_1)^{ab}_{dc}=\delta_{ac}-\frac{1}{\d_v} \sqrt{\frac{\d_a \d_c}{\d_b \d_d}}\delta_{bd}\,.
\end{equation}
The coefficients of the amplitude in the projector basis \eqref{eq:Sproj} are
\begin{equation}
    A_0=Z(\theta)\sinh\(\frac{i\pi+\theta}{n}\)\,, \quad A_1=Z(\theta)\sinh\(\frac{i\pi-\theta}{n}\)\,,
\end{equation}
which satisfy $A_\chi(\theta)A_\chi(-\theta)=1$ as one can check from \eqref{eq:Zdef} and reproduce \eqref{eq:SRSOS}.
\subsection{\texorpdfstring{Sine-Gordon model and SG$(\beta,2)$}{Sine-Gordon model and SG(beta,2)}}In the main text, we discussed the perturbed $SU(2)_1$ WZW model as an example of theories with $\mathbb{Z}_2$ anomaly. It has an alternative description as a compact boson at radius $R=\sqrt{2}$ with the deformation $\cos (2X/R)$. Here we propose the exact $S$-matrix for such theories at arbitrary $R$ (called $SG(\beta,2)$ in \cite{Klassen:1992eq,Klassen:1992qy}), which obeys the same modified crossing rule.

First let us contrast $SG(\beta,2)$ with the standard sine-Gordon model, which is a compact boson perturbed by $\cos (X/R)$. The sine-Gordon theory has a single vacuum and the coupling $\beta$ is determined by the radius as
\be
\beta^2=\frac{4\pi}{R^2}\,.
\ee
The S-matrix of solitons is given by \cite{Zamolodchikov:1978xm}
\be
S_{SG}=\(\begin{array}{cccc}S_0&0&0&0\\0&S_{T}&S_{R}&0\\0&S_R&S_T&0\\0&0&0&S_0\end{array}\)\,,
\ee
where $S_T$ and $S_R$ are transmission and reflection amplitudes and read
\be\label{eq:STSR}
S_T=\frac{\sinh(\pi\theta/\xi)\, S_0}{\sinh(\pi(i\pi - \theta)/\xi)}\,, \quad S_R=\frac{i\sin(\pi^2/\xi)\, S_0}{\sinh(\pi(i\pi - \theta)/\xi)}\,,
\ee
and
\be\label{eq:SGS0}
S_0=-\exp\left(-i \int_0^{\infty} \frac{dt}{t} \frac{\sinh\left(t(\pi - \xi)\right)}{\sinh(\xi t) \cosh(\pi t)} \sin(2\theta t)\right)\,.
\ee
where $\xi$ is given by
\be
\xi=\frac{\beta^2}{8}\frac{1}{1-\frac{\beta^2}{8\pi}}=\frac{\pi}{2R^2-1}\,.
\ee

We now consider $SG(\beta,2)$, whose S-matrix has not been identified in the literature. The dynamics of this theory is locally the same as sine-Gordon theory, up to a redefinition of the coupling
\be\label{eq:redefC}
\tilde{\beta}^2=\frac{16\pi }{R^2}\,.
\ee
However there are also  crucial differences: 1.~the theory has two vacua 2.~the theory has $\mathbb{Z}_2$ anomaly. We propose the following S-matrix compatible with the aforementioned properties:
\be
S_{SG(\beta,2)}=S_{\text{RSOS}_2}\otimes \check{S}_{SG}\,,
\ee
where $\check{S}_{SG}$ describes scattering among Kramers pairs and reads
\be
\check{S}_{SG}=\(\begin{array}{cccc}S_0&0&0&0\\0&-S_{T}&S_{R}&0\\0&S_R&-S_T&0\\0&0&0&S_0\end{array}\)\,.
\ee
where $S_0$, $S_{T}$ and $S_{R}$ are the same as \eqref{eq:STSR} and \eqref{eq:SGS0} except that the coupling is redefined as \eqref{eq:redefC}, i.e.
\be
\xi=\frac{\tilde{\beta}^2}{8}\frac{1}{1-\frac{\tilde{\beta}^2}{8\pi}}=\frac{\pi}{\frac{R^2}{2}-1}\,.
\ee

The S-matrix $\check{S}_{SG}$ obeys the same modified crossing rule as the perturbed WZW model \eqref{eq:modifiedcrossZ2}. It can be derived formally from the standard sine-Gordon S-matrix by expressing it in terms of the R-matrix of $U_q(sl_2)$ and reversing the sign of $q$ \cite{bernard1991quantum}. Despite being often dubbed as an ``alternative" S-matrix for the standard sine-Gordon model \cite{bernard1991quantum} due to structural similarities between $U_q (sl_2)$ and $U_{-q}(sl_2)$, we propose that it describes a distinct model, $SG(\beta,2)$, with modified crossing as a consequence of the $\mathbb{Z}_2$ anomaly. 

There exists a one-parameter family of models generalizing $SG(\beta, 2)$: perturbed compact bosons by $\cos (k X/R)$ with $k\in \mathbb{N}_{>0}$, called $SG(\beta,k)$ in \cite{Klassen:1992eq,Klassen:1992qy}. Their local dynamics mirror those of the sine-Gordon theory, albeit with a coupling redefinition $\beta^2=4\pi k^2/R^2$. The preserved $\bZ_k$ symmetry is anomalous and the IR theory has $k$ distinct vacua.
Exploring the S-matrices of these theories and understanding their relation to symmetries and anomalies would be interesting.

\end{appendices}

\end{document}